\documentclass[preprint,12pt]{elsarticle}

\usepackage[utf8]{inputenc}
\usepackage[T1]{fontenc}
\usepackage{amsmath,amssymb,amsfonts}
\usepackage[a4paper,margin=2.5cm]{geometry}
\usepackage{microtype}
\usepackage[colorlinks=true,linkcolor=blue,citecolor=blue,urlcolor=blue]{hyperref}

\biboptions{sort&compress}
\journal{Annals of Physics}

\begin{document}
	
	\begin{frontmatter}
		
		\title{The inverted harmonic oscillator in Segal--Bargmann space:\\ Time-asymmetric propagator reconstruction\\ and bi-coherent states}
		
		\author[sxu]{Hongze Ding}
		\ead{dinghongze@sxu.edu.cn}
		
		\author[sxu]{Ni Liu\corref{cor1}}
		\ead{liuni2011520@sxu.edu.cn}
		
		\author[sxu]{J.-Q. Liang\corref{cor2}}
		\ead{jqliang@sxu.edu.cn}
		
		\cortext[cor1]{Corresponding author.}
		\cortext[cor2]{Principal corresponding author.}
		
		\address[sxu]{Institute of Theoretical Physics, State Key Laboratory of Quantum Optics and Quantum Optics Devices, Collaborative Innovation Center of Extreme Optics, Shanxi University, Taiyuan 030006, Shanxi, People's Republic of China}
		
		\begin{abstract}
			In this paper, we investigate the real-time dynamics and resonance
			quantization of the inverted harmonic oscillator (IHO), a formally Hermitian
			system whose potential is unbounded from below. Although its Hamiltonian is
			essentially self-adjoint on $L^2(\mathbb{R})$ with a continuous real
			scattering spectrum, outside $L^2(\mathbb{R})$ it admits unstable discrete
			resonance eigenstates carrying purely imaginary eigenvalues $\pm
			i\omega\hbar(n + \frac{1}{2})$ that govern exponential decay and growth.
			Within the imaginary-frequency boson framework, these dual ``bra'' and
			``ket'' eigenstates are realized in coordinate space as complex Hermite
			polynomials biorthonormalized under the Fresnel Gaussian measure. As our
			central result, we derive the exact real-time Feynman propagator directly
			from the path integral via a real-time adaptation of Moriconi's
			coordinate-time transformation and prove analytically, using Mehler's
			formula, that the full unitary propagator is reconstructed term-by-term
			exclusively from the discrete imaginary-eigenvalue ladder with no continuum
			contribution. Remarkably, this unique unitary kernel splits into two
			mutually exclusive discrete expansions with opposite temporal convergence
			domains: the decaying Gamow--Siegert branch converges solely for forward
			propagation ($T>0$), whereas the growing anti-resonant branch converges
			solely for backward propagation ($T<0$). To clarify the operator-theoretic
			origin of this time-asymmetric splitting and real-axis doubling, we further
			develop a complex-scaled Segal--Bargmann representation on $\mathcal{F}(%
			\mathbb{C})$, where the Hamiltonian reduces to the Euler operator $\pm
			i\omega\hbar(z\,d/dz + \frac{1}{2})$. Analyticity at $z=0$ yields a single
			orthonormal monomial basis $z^n/\sqrt{n!}$ (recovering the dual complex
			Hermite states upon inversion), while time evolution reduces to one-sided
			contraction semigroups of complex dilatations on $\mathcal{F}(\mathbb{C})$
			for $T>0$ and $T<0$ mirroring the two Mehler branches. Finally, we construct
			the imaginary-frequency coherent states ($e^{\alpha z}$ in $\mathcal{F}(%
			\mathbb{C})$), verify $\Delta\mathrm{x}\,\Delta\mathrm{p} = \hbar/2$ for
			their conjugate imaginary biorthogonal second cumulants, and show that their
			dual Heisenberg expectation values match the exact classical hyperbolic
			trajectories.
		\end{abstract}
		
		\begin{keyword}
			inverted harmonic oscillator \sep Feynman propagator \sep resonance expansion \sep Mehler formula \sep Segal--Bargmann space \sep complex scaling \sep bi-coherent states
		\end{keyword}
		
	\end{frontmatter}
	
	\section{Introduction}
	
	It is well known in conventional quantum mechanics that a Hermitian
	Hamiltonian on the Hilbert space $L^2(\mathbb{R})$ guarantees a real energy
	spectrum and the conservation of probability, whereas non-Hermitian
	Hamiltonians usually describe open systems with complex energy spectra. Over
	the past two decades, however, Bender and collaborators demonstrated that
	formal Hermiticity is not necessary for a Hamiltonian to possess real
	eigenvalues~\cite{Bender1998,Bender1999,Bender2002}. In particular,
	non-Hermitian Hamiltonians can exhibit entirely real spectra if they respect
	parity-time ($\mathcal{PT}$) reflection symmetry~\cite%
	{Bender1998,Bender1999,Bender2002,Bender2004,Bender2007,Bender2013,Bender2018,Christodoulides2018,Gu2022a,Gu2022b,Liu2022}%
	. More generally, Mostafazadeh showed that non-Hermitian Hamiltonians with
	real spectra are characterized by pseudo-Hermiticity~\cite%
	{Mostafazadeh2002a,Mostafazadeh2002b,Mostafazadeh2007,Mostafazadeh2010}, a
	framework that is not confined to $\mathcal{PT}$ symmetry; indeed, $\mathcal{%
		PT}$-asymmetric non-Hermitian Hamiltonians can also possess purely real
	spectra~\cite{Amaaouche2022,Liu2023,Luo2024}.
	
	According to standard spectral theory, a self-adjoint operator on $L^2(%
	\mathbb{R})$ has a purely real spectrum~\cite%
	{Reed1980,Mostafazadeh2010,Scholtz1992}. However, formal Hermiticity of a
	Hamiltonian is neither necessary nor sufficient to ensure real eigenvalues.
	While the non-necessity has been extensively studied in recent years through
	non-Hermitian operators with real spectra, the insufficiency is much less
	appreciated. In fact, formal Hermiticity guarantees real discrete
	eigenvalues only under the implicit assumption that the potential is bounded
	from below, so that boundary terms arising from integration by parts vanish
	for square-integrable states~\cite{Reed1980,Mostafazadeh2010,Scholtz1992}.
	For a conservative system whose potential is unbounded from below, a
	formally Hermitian Hamiltonian can admit generalized eigenstates outside $%
	L^2(\mathbb{R})$ with complex eigenvalues. For example, Gamow--Siegert
	states with complex eigenvalues were introduced long ago to describe $\alpha$%
	-decay in nuclear physics~\cite{Gamow1928,Siegert1939,Civitarese2004}.
	Complex spectra have also been studied theoretically for inverted
	double-well potentials via the instanton method~\cite{Liang1992,Liang1994},
	revealed in parabolic barrier scattering~\cite{Palma2004}, found in
	minisuperspace models of an expanding universe with a cosmological constant~%
	\cite{Liang1999}, and linked to black hole quasinormal modes near the
	maximum of an effective potential barrier~\cite{Ferrari1984}. The inverted
	harmonic oscillator (IHO)~\cite{Barton1986,Subramanyan2021} is the
	prototypical extreme case of such barrier-top instability: although $%
	\widehat{H}$ is essentially self-adjoint on $L^2(\mathbb{R})$ with a
	continuous real scattering spectrum, its unstable resonance eigenstates
	outside $L^2(\mathbb{R})$ carry purely imaginary eigenvalues $\pm
	i\omega\hbar(n + \frac{1}{2})$~\cite%
	{Shimbori2000a,Shimbori2000b,Chruscinski2004,Chruscinski2006,Baskoutas1993,Bermudez2013,Bagarello2022,Bagarello2026}%
	. Unlike bound states with real eigenvalues, the probability density of an
	individual eigenstate is no longer conserved, and the imaginary part of the
	energy eigenvalue characterizes the decay or growth rate of the unstable
	state~\cite{Liang1992,Liang1994}. In particular, understanding how the exact
	unitary real-time Feynman propagator---which on $L^2(\mathbb{R})$ is
	conventionally governed by the continuous real scattering spectrum---can be
	reconstructed from these non-$L^2(\mathbb{R})$ discrete resonance states,
	and how time asymmetry emerges between forward and backward propagation
	within a self-consistent algebraic and phase-space framework, remains a
	foundational problem in quantum mechanics.
	
	In recent work~\cite{Liu2024,Liu2026}, we solved the eigenvalue problem of
	the IHO algebraically by introducing imaginary-frequency raising and
	lowering boson operators obtained via the analytic continuation $\Omega =
	i\omega$ of the normal harmonic oscillator (see also the distributional $%
	\mathcal{D}$-pseudo-boson formulation~\cite{Bagarello2022,Bagarello2026} and
	pseudo-Hermitian metric constructions~\cite{Zerimeche2023,Amaouche2022b}).
	Because the resulting boson number operator is non-Hermitian while its
	integer eigenvalues remain real, dual sets of ``bra'' and ``ket''
	eigenstates are naturally required in coordinate space, where they are
	realized as spatially delocalized complex Hermite polynomials
	biorthonormalized under the Fresnel Gaussian measure natural to Feynman path
	integrals. The primary focus of the present paper is to establish the exact
	equivalence and time-asymmetric spectral reconstruction of the real-time
	Feynman propagator of the IHO in terms of these discrete resonance
	eigenstates. We first derive the exact closed-form propagator~\cite%
	{Barton1986} directly from the inverted-oscillator path integral by adapting
	Moriconi's coordinate-time transformation~\cite{Moriconi2004} to real time,
	providing an independent path-integral benchmark. Inserting the dual sets of
	complex Hermite eigenstates and applying Mehler's summation formula~\cite%
	{Lebedev1963,Liang1991}, we prove analytically that the discrete
	imaginary-eigenvalue ladder alone rebuilds the full unitary propagator term
	by term, with no explicit contribution from the continuous real scattering
	spectrum. Crucially, this discrete spectral representation splits into two
	disjoint, time-asymmetric branches with mutually exclusive domains of
	convergence: the decaying Gamow--Siegert expansion converges solely for
	forward time evolution ($T>0$), whereas the growing anti-resonant expansion
	converges solely for backward time evolution ($T<0$), with each branch
	summing to the exact same unitary kernel in its respective time domain.
	
	To clarify the Hilbert-space and operator-theoretic origin of both this
	forward\slash backward propagator branch splitting and the real-axis
	biorthogonal doubling, we further develop a complex-scaled Segal--Bargmann
	representation~\cite%
	{Bargmann1961,Bargmann1967,Segal1963,Hall2000,Vourdas2006} of the
	imaginary-frequency boson algebra. By combining the Bargmann transform with
	the complex coordinate rotation $x \to e^{\mp i\pi/4}u$~\cite%
	{Chruscinski2004,Aguilar1971,Balslev1971,Moiseyev2011,Moshinsky1971,Wolf1974}%
	, the rotated IHO Hamiltonians become skew-adjoint normal operators on $L^2(%
	\mathbb{R}_u)$ and map onto the holomorphic space $\mathcal{F}(\mathbb{C})$
	of entire functions as the first-order Euler operators $\pm
	i\omega\hbar(z\,d/dz + \frac{1}{2})$. Single-valued analyticity at $z=0$
	directly quantizes the imaginary spectrum $\pm i\omega\hbar(n + \frac{1}{2})$
	and yields the single orthonormal monomial basis $z^n/\sqrt{n!}$ in each
	sector (cf.~\cite{Ali2015}), while the inverse Fresnel--Bargmann transforms
	recover the exact complex Hermite wave functions and show their Fresnel
	biorthogonal pairing to be the deformed unitary Bargmann inner product. In $%
	\mathcal{F}(\mathbb{C})$, moreover, the time-evolution operators act simply
	as weighted complex dilatations that define bounded contraction semigroups
	exclusively for $T \ge 0$ (decaying branch) and $T \le 0$ (growing branch),
	directly explaining the one-sided convergence domains of the two Mehler
	propagator expansions. Finally, we construct the imaginary-frequency boson
	coherent states as eigenstates of the lowering operators (corresponding to $%
	e^{\alpha z}$ in $\mathcal{F}(\mathbb{C})$), verify the minimum uncertainty
	relation $\Delta\mathrm{x}\,\Delta\mathrm{p} = \hbar/2$ for their conjugate
	imaginary biorthogonal second cumulants, and demonstrate that their dual
	Heisenberg expectation values agree precisely with the classical hyperbolic
	equations of motion for arbitrary real initial conditions $(x_0, p_0)$,
	establishing exact quantum--classical correspondence in the
	imaginary-eigenvalue system.
	
	The remainder of this paper is organized as follows. In section~\ref%
	{sec:prelim}, the dual sets of eigenstates are obtained by means of the
	imaginary-frequency boson algebraic method for the IHO Hamiltonian (section~%
	\ref{sec:dual}), and the normalization of the wave functions together with
	the complex Hermite polynomials is presented under the Fresnel Gaussian
	measure (section~\ref{sec:hermite}). In section~\ref{sec:bargmann}, we
	establish the complex-scaled Segal--Bargmann representation as the analytic
	framework underlying the resonance sectors, derive the orthonormal monomial
	basis from entire-function regularity, recover the complex Hermite wave
	functions via the inverse Bargmann transform, and obtain the one-sided
	time-evolution semigroups in $\mathcal{F}(\mathbb{C})$. Section~\ref%
	{sec:propagator} presents our central dynamical results: we derive the exact
	real-time Feynman propagator via Moriconi's path-integral transformation and
	prove via Mehler's formula its strict, time-asymmetric reconstruction from
	the dual sets of discrete eigenstates in exact correspondence with the
	Bargmann semigroup domains. The imaginary-frequency boson coherent states
	and their minimum uncertainty relation are demonstrated in section~\ref%
	{sec:coherent}, followed by the dynamical quantum--classical correspondence
	in section~\ref{sec:decay}. Finally, conclusions and discussions are
	presented in section~\ref{sec:conclusion}.
	
	\section{Imaginary-frequency boson algebra and complex Hermite wave functions%
	}
	
	\label{sec:prelim}
	
	\subsection{Brief review of the algebraic formulation and dual-set
		eigenstates}
	
	\label{sec:dual}
	
	To make the subsequent path-integral, Bargmann-space, and coherent-state
	analyses self-contained, we briefly recall the imaginary-frequency boson
	formulation of the IHO~\cite{Liu2024,Liu2026}. We consider a particle of
	unit mass ($m=1$) subject to an inverted harmonic-oscillator potential
	described by the Hamiltonian  
	\begin{equation}
		\widehat{H} = \frac{\widehat{\mathrm{p}}^2}{2} - \frac{1}{2}\omega^2\widehat{%
			\mathrm{x}}^2,  \label{eq:1}
	\end{equation}
	where $\omega > 0$. Classically, a particle released from the potential
	maximum $\mathrm{x}(0) = 0$ with initial speed $v$ follows the runaway
	hyperbolic trajectory  
	\begin{equation}
		\mathrm{x}(t) = \pm \frac{v}{\omega}\sinh\omega t,  \label{eq:2}
	\end{equation}
	reducing to an unstable static sphaleron-like configuration when $v=0$~\cite%
	{MullerKirsten2006,Liang2023}. Quantum mechanically, although $\widehat{H}$
	is essentially self-adjoint on $L^2(\mathbb{R})$ with a continuous real
	spectrum, the generalized eigenstates governing exponential decay and growth
	reside outside $L^2(\mathbb{R})$ and carry complex conjugate pure imaginary
	eigenvalues~\cite%
	{Reed1980,Mostafazadeh2010,Scholtz1992,Bagarello2022,Shimbori2000a,Chruscinski2004}%
	.
	
	To understand the algebraic origin of these states~\cite{Liu2024,Liu2026},
	consider the standard harmonic oscillator of unit mass $\widehat{H}%
	_{\mathrm{HO}}=\frac{1}{2}\widehat{\mathrm{p}}^{2}+\frac{1}{2}\Omega ^{2}%
	\widehat{\mathrm{x}}^{2}$ with positive frequency $\Omega >0$, for which $%
	\widehat{H}_{\mathrm{HO}}=\hbar \Omega (\widehat{a}^{\dagger }(\Omega )%
	\widehat{a}(\Omega )+\frac{1}{2})$ in terms of the usual ladder operators $%
	\widehat{a}(\Omega )=\sqrt{\Omega /(2\hbar )}(\widehat{\mathrm{x}}+i\widehat{%
		\mathrm{p}}/\Omega )$ and $\widehat{a}^{\dagger }(\Omega )=\sqrt{\Omega
		/(2\hbar )}(\widehat{\mathrm{x}}-i\widehat{\mathrm{p}}/\Omega )$. Performing
	the analytic continuation 
	\begin{equation}
		\Omega =i\omega \qquad (\omega >0),  \label{eq:analytic_cont}
	\end{equation}%
	maps $\widehat{H}_{\mathrm{HO}}$ directly onto the IHO Hamiltonian~%
	\eqref{eq:1}. Introducing the dimensionless canonical operators $\widehat{x}=%
	\sqrt{\omega /\hbar }\,\widehat{\mathrm{x}}$ and $\widehat{p}=\widehat{%
		\mathrm{p}}/\sqrt{\omega \hbar }$ satisfying $[\widehat{x},\widehat{p}]=i$,
	equation~\eqref{eq:1} reads 
	\begin{equation}
		\widehat{H}=i\omega \hbar \left[ \frac{\widehat{p}^{2}}{2i}+\frac{1}{2}i%
		\widehat{x}^{2}\right] .  \label{eq:8}
	\end{equation}%
	Under the continuation~\eqref{eq:analytic_cont} (with principal root $\sqrt{i%
	}=e^{i\pi /4}$), the harmonic-oscillator ladder operators become the
	imaginary-frequency boson operators~\cite{Liu2024,Liu2026} 
	\begin{equation}
		\widehat{a}_{-}=\sqrt{\frac{i}{2}}(\widehat{x}+\widehat{p}),\qquad \widehat{a%
		}_{+}=\sqrt{\frac{i}{2}}(\widehat{x}-\widehat{p}).  \label{eq:9}
	\end{equation}%
	Using $(\sqrt{i/2})^{\ast }=-i\sqrt{i/2}$, one verifies that they satisfy
	the canonical bosonic commutation relation together with modified Hermitian
	conjugation properties on $L^{2}(\mathbb{R})$: 
	\begin{equation}
		[\widehat{a}_{-},\widehat{a}_{+}]=1,\qquad \widehat{a}_{-}^{\dagger
		}=-i\widehat{a}_{-},\qquad \widehat{a}_{+}^{\dagger }=-i\widehat{a}_{+}.
		\label{eq:10}
	\end{equation}%
	Because $[\widehat{a}_{-},\widehat{a}_{+}]=1$ with $\widehat{a}_{+}\neq 
	\widehat{a}_{-}^{\dagger }$, the pair $(\widehat{a}_{-},\widehat{a}_{+})$
	constitutes a concrete realization of $\mathcal{D}$-pseudo-bosons~\cite%
	{Bagarello2022,Shimbori2000b} and generates an $\mathfrak{su}(1,1)$
	dynamical algebra via $\widehat{S}_{z}=\frac{1}{2}(\widehat{a}_{+}\widehat{a}%
	_{-}+\frac{1}{2})$ and $\widehat{S}_{\pm }=\frac{1}{2}\widehat{a}_{\pm }^{2}$%
	~\cite{Liu2026,Bagarello2026}. Defining the non-Hermitian number operator $%
	\widehat{n}=\widehat{a}_{+}\widehat{a}_{-}$, which obeys 
	\begin{equation}
		\widehat{n}^{\dagger }=-\widehat{a}_{-}\widehat{a}_{+}=-(\widehat{n}+1),
		\label{eq:12}
	\end{equation}%
	the Hamiltonian~\eqref{eq:8} assumes the factorized form 
	\begin{equation}
		\widehat{H}=i\omega \hbar \left( \widehat{n}+\frac{1}{2}\right) =\widehat{H}%
		^{\dagger },  \label{eq:11}
	\end{equation}%
	where the formal Hermiticity $\widehat{H}^{\dagger }=\widehat{H}$ follows
	directly from equation~\eqref{eq:12}: $-i\omega \hbar (\widehat{n}^{\dagger
	}+\frac{1}{2})=-i\omega \hbar (-\widehat{n}-\frac{1}{2})=\widehat{H}$.
	
	For non-negative integer eigenvalues $n = 0,1,2,\dots$, the pseudo-Hermitian
	number operator $\widehat{n}$ admits dual sets of ``ket'' ($|n\rangle_r$)
	and ``bra'' ($|n\rangle_l$) eigenstates~\cite{Liu2024,Liu2026},  
	\begin{equation}
		\begin{aligned} \widehat{n}|n\rangle_r &= n|n\rangle_r, & {}_r\langle
			n|\widehat{n}^\dagger &= n\,{}_r\langle n|, \\
			\widehat{n}^\dagger|n\rangle_l &= n|n\rangle_l, & {}_l\langle n|\widehat{n}
			&= n\,{}_l\langle n|, \end{aligned}  \label{eq:13}
	\end{equation}
	normalized by the biorthogonal relation  
	\begin{equation}
		{}_l\langle n|m\rangle_r = \delta_{nm}.  \label{eq:14}
	\end{equation}
	Using $\widehat{n}|n\rangle_l = -(\widehat{n}^\dagger + 1)|n\rangle_l =
	-(n+1)|n\rangle_l$, the right and left eigenvalue equations for $\widehat{H}$
	yield complex conjugate pure imaginary eigenvalues:  
	\begin{equation}
		\widehat{H}|n\rangle_r = i\omega\hbar\left(n + \frac{1}{2}%
		\right)|n\rangle_r, \qquad \widehat{H}|n\rangle_l = -i\omega\hbar\left(n + 
		\frac{1}{2}\right)|n\rangle_l,  \label{eq:15}
	\end{equation}
	\begin{equation}
		{}_r\langle n|\widehat{H} = -i\omega\hbar\left(n + \frac{1}{2}%
		\right)\,{}_r\langle n|, \qquad {}_l\langle n|\widehat{H} =
		i\omega\hbar\left(n + \frac{1}{2}\right)\,{}_l\langle n|.  \label{eq:16}
	\end{equation}
	Accordingly, the dual expectation values ${}_l\langle n|\widehat{H}%
	|n\rangle_r = [{}_r\langle n|\widehat{H}|n\rangle_l]^* = i\omega\hbar(n + 
	\frac{1}{2})$ are finite, whereas single-set overlaps ${}_i\langle
	n|n\rangle_i$ ($i=r,l$) diverge outside $L^2(\mathbb{R})$.
	
	From the commutation relation~\eqref{eq:10} and biorthogonality~\eqref{eq:14}%
	, $\widehat{a}_-$ and $\widehat{a}_+$ act respectively as lowering and
	raising operators on $|n\rangle_r$, while their roles are interchanged on $%
	|n\rangle_l$~\cite{Liu2024,Liu2026}:  
	\begin{equation}
		\begin{aligned} \widehat{a}_-|n\rangle_r &= \sqrt{n}\,|n-1\rangle_r, &
			\widehat{a}_+|n-1\rangle_r &= \sqrt{n}\,|n\rangle_r, \\
			\widehat{a}_+|n\rangle_l &= i\sqrt{n}\,|n-1\rangle_l, &
			\widehat{a}_-|n-1\rangle_l &= i\sqrt{n}\,|n\rangle_l, \end{aligned}
		\label{eq:19}
	\end{equation}
	subject to the dual vacuum conditions  
	\begin{equation}
		\widehat{a}_-|0\rangle_r = 0, \qquad \widehat{a}_+|0\rangle_l = 0,
		\label{eq:21}
	\end{equation}
	from which the $n$-th dual Fock states are generated as  
	\begin{equation}
		|n\rangle_r = \frac{(\widehat{a}_+)^n}{\sqrt{n!}}|0\rangle_r, \qquad
		|n\rangle_l = \frac{(\widehat{a}_-)^n}{i^n\sqrt{n!}}|0\rangle_l.
		\label{eq:23}
	\end{equation}
	
	\subsection{Complex Hermite polynomial wave functions}
	
	\label{sec:hermite}
	
	In the dimensionless coordinate representation ($\widehat{p} = -i\frac{d}{dx}
	$), the ladder operators~\eqref{eq:9} act as the first-order differential
	operators  
	\begin{equation}
		\widehat{a}_- = \sqrt{\frac{i}{2}}\left(x - i\frac{d}{dx}\right), \qquad 
		\widehat{a}_+ = \sqrt{\frac{i}{2}}\left(x + i\frac{d}{dx}\right).
		\label{eq:ladder_diff}
	\end{equation}
	The vacuum conditions~\eqref{eq:21} become $\left(x \mp i\frac{d}{dx}%
	\right)\psi_0^{r,l}(x) = 0$, admitting the complex Gaussian solutions $%
	\psi_0^r(x) = N_r e^{-ix^2/2}$ and $\psi_0^l(x) = N_l e^{ix^2/2}$~\cite%
	{Liu2026}. Under the Fresnel Gaussian integration measure natural to Feynman
	path integrals~\cite{Liang2023,Feynman1965,Liu2026}, the biorthogonal
	condition $\int_{-\infty}^{\infty} [\psi_0^l(x)]^*\psi_0^r(x)\,dx = N_l^*
	N_r \int_{-\infty}^{\infty} e^{-ix^2}\,dx = N_l^* N_r \sqrt{\pi/i} = 1$
	yields the symmetric normalization constants  
	\begin{equation}
		N_r = \left(\frac{i}{\pi}\right)^{\frac{1}{4}}, \qquad N_l = \left(\frac{-i}{%
			\pi}\right)^{\frac{1}{4}}.  \label{eq:31}
	\end{equation}
	By contrast, the single-set probability density $|\psi_0^{r,l}(x)|^2 =
	\pi^{-1/2}$ is constant across $\mathbb{R}$, confirming that these states
	lie outside $L^2(\mathbb{R})$~\cite{Bagarello2022}.
	
	Acting with the raising operators in equation~\eqref{eq:23} on $%
	\psi_0^{r,l}(x)$ produces the $n$-th coordinate wave functions  
	\begin{equation}
		\begin{aligned}
			\psi_n^r(x) &= \frac{N_r}{\sqrt{n!}}\left(\sqrt{\frac{i}{2}}\right)^n \left(x + i\frac{d}{dx}\right)^n e^{-i\frac{x^2}{2}}, \\
			\psi_n^l(x) &= \frac{N_l}{i^n\sqrt{n!}}\left(\sqrt{\frac{i}{2}}\right)^n \left(x - i\frac{d}{dx}\right)^n e^{i\frac{x^2}{2}}.
		\end{aligned}
		\label{eq:29}
	\end{equation}
	To cast the differential-operator representation~\eqref{eq:29} into closed
	algebraic form suitable for the Bargmann and Mehler-kernel calculations
	below, we define the complex Hermite polynomials~\cite%
	{Lebedev1963,Chruscinski2004}  
	\begin{equation}
		\begin{aligned} H_n(x) &= e^{i\frac{x^2}{2}}\left(x + i\frac{d}{dx}\right)^n
			e^{-i\frac{x^2}{2}} =
			(\sqrt{i})^{-n}\mathcal{H}_n^{\mathrm{std}}(\sqrt{i}\,x), \\ H_n^*(x) &=
			e^{-i\frac{x^2}{2}}\left(x - i\frac{d}{dx}\right)^n e^{i\frac{x^2}{2}} =
			(\sqrt{-i})^{-n}\mathcal{H}_n^{\mathrm{std}}(\sqrt{-i}\,x), \end{aligned}
		\label{eq:complex_H}
	\end{equation}
	where $\mathcal{H}_n^{\mathrm{std}}(w) = (-1)^n e^{w^2}\frac{d^n}{dw^n}%
	e^{-w^2}$ denotes the standard Hermite polynomial analytically continued to $%
	w\in\mathbb{C}$. The dual eigenstates then take the compact form  
	\begin{equation}
		\begin{aligned}
			\psi_n^r(x) &= \frac{1}{\sqrt{n!}}\left(\sqrt{\frac{i}{2}}\right)^n \left(\frac{i}{\pi}\right)^{\frac{1}{4}} e^{-i\frac{x^2}{2}} H_n(x), \\
			\psi_n^l(x) &= \frac{1}{i^n\sqrt{n!}}\left(\sqrt{\frac{i}{2}}\right)^n \left(\frac{-i}{\pi}\right)^{\frac{1}{4}} e^{i\frac{x^2}{2}} H_n^*(x),
		\end{aligned}
	\end{equation}
	which satisfy the conjugation symmetry $\psi_n^l(x) = [\psi_n^r(x)]^*$. For
	all $n,m\ge 0$, the coordinate biorthogonality $\int_{-\infty}^{\infty}
	[\psi_m^l(x)]^*\psi_n^r(x)\,dx = \delta_{mn}$ holds under Fresnel
	regularization (introducing $e^{-\epsilon x^2}$ with $\epsilon\to 0^+$ or
	rotating the integration contour via $x = e^{-i\pi/4}u$), in full agreement
	with the rigged Hilbert space formulation~\cite%
	{Bagarello2022,Shimbori2000a,Chruscinski2004}. Restoring the physical
	coordinate $\mathrm{x} = \sqrt{\hbar/\omega}\,x$, the normalized dual
	coordinate eigenfunctions read  
	\begin{align}
		\psi_n^r(\mathrm{x}) &= \frac{1}{\sqrt{n!}}\left(\sqrt{\frac{i}{2}}\right)^n
		\left(\frac{i\omega}{\pi\hbar}\right)^{\frac{1}{4}} e^{-i\frac{\omega\mathrm{%
					x}^2}{2\hbar}} H_n\!\left(\sqrt{\frac{\omega}{\hbar}}\,\mathrm{x}\right),
		\label{eq:32} \\
		\psi_n^l(\mathrm{x}) &= \frac{1}{i^n\sqrt{n!}}\left(\sqrt{\frac{i}{2}}%
		\right)^n \left(\frac{-i\omega}{\pi\hbar}\right)^{\frac{1}{4}} e^{i\frac{%
				\omega\mathrm{x}^2}{2\hbar}} H_n^*\!\left(\sqrt{\frac{\omega}{\hbar}}\,%
		\mathrm{x}\right).  \label{eq:33}
	\end{align}
	
	\section{Segal--Bargmann formulation and orthonormal resonance basis}
	
	\label{sec:bargmann}
	
	In section~\ref{sec:prelim}, staying directly on the real coordinate axis $%
	\mathbb{R}$ forced us to introduce a biorthogonal dual pair $(\psi_n^r,
	\psi_n^l)$ outside $L^2(\mathbb{R})$ because $\widehat{a}_+ \neq \widehat{a}%
	_-^\dagger$ in $L^2(\mathbb{R})$. Before turning to our central analysis of
	the Feynman propagator in section~\ref{sec:propagator}, we show in this
	section that passing via complex scaling to the holomorphic Segal--Bargmann
	space $\mathcal{F}(\mathbb{C})$ eliminates the need for a biorthogonal dual
	basis within each sector's representation space, yields a single orthonormal
	Fock basis of entire functions (recovering equations~\eqref{eq:32}--%
	\eqref{eq:33} upon inversion), and establishes the one-sided
	contraction-semigroup kernels governing forward ($T>0$) and backward ($T<0$)
	time evolution.
	
	\subsection{Complex scaling, normal-operator structure, and Stone--von
		Neumann equivalence}
	
	Before invoking the Stone--von Neumann theorem~\cite%
	{Stone1930,vonNeumann1931,Folland1989,Hall2013}, an important spectral
	distinction must be made explicit. On the real-axis Hilbert space $L^2(%
	\mathbb{R})$, the Hamiltonian~\eqref{eq:1} is self-adjoint with a continuous
	real spectrum $(-\infty,\infty)$, so no unitary transformation acting
	directly from $L^2(\mathbb{R})$ can convert $\widehat{H}$ into a
	skew-adjoint operator with discrete pure imaginary eigenvalues. Instead, the
	resonance sectors are uncovered by a complex coordinate scaling (complex
	dilation)~\cite{Aguilar1971,Balslev1971,Moiseyev2011,Chruscinski2004}.
	
	For the incoming (or growing-eigenvalue, anti-resonant) sector, consider the
	rotated contour $\Gamma_+ = e^{-i\pi/4}\mathbb{R}$ parameterized by $x =
	e^{-i\pi/4}u$ with $u = \sqrt{i}\,x \in \mathbb{R}$ (so $\frac{d}{dx} = 
	\sqrt{i}\,\frac{d}{du}$). Substituting this rotation into equation~%
	\eqref{eq:ladder_diff} turns the imaginary-frequency operators $\widehat{a}_-
	$ and $\widehat{a}_+$ into standard harmonic-oscillator ladder operators on $%
	L^2(\mathbb{R}_u)$:  
	\begin{equation}
		\widehat{b}_r := \widehat{a}_- = \frac{1}{\sqrt{2}}\left(u + \frac{d}{du}%
		\right), \qquad \widehat{b}_r^\dagger = \widehat{a}_+ = \frac{1}{\sqrt{2}}%
		\left(u - \frac{d}{du}\right), \qquad [\widehat{b}_r, \widehat{b}_r^\dagger]
		= 1,  \label{eq:b_rotated}
	\end{equation}
	which are genuine mutual Hilbert-space adjoints on $L^2(\mathbb{R}_u)$.
	Correspondingly, the complex-scaled Hamiltonian takes the form  
	\begin{equation}
		\widehat{H}_+ = i\omega\hbar\left(\widehat{b}_r^\dagger\widehat{b}_r + \frac{%
			1}{2}\right) = i\omega\hbar\left(-\frac{1}{2}\frac{d^2}{du^2} + \frac{1}{2}%
		u^2\right),  \label{eq:H_plus}
	\end{equation}
	in agreement with the complex-dilation result of~\cite{Chruscinski2004}, and
	satisfies $\widehat{H}_+^\dagger = -\widehat{H}_+$ on $L^2(\mathbb{R}_u)$.
	Consequently, $\widehat{H}_+$ is skew-adjoint and therefore a \emph{normal
		operator} on $L^2(\mathbb{R}_u)$:  
	\begin{equation}
		[\widehat{H}_+, \widehat{H}_+^\dagger] = 0.  \label{eq:H_normal}
	\end{equation}
	Since $\widehat{H}_+ = i\omega\hbar(\widehat{b}_r^\dagger\widehat{b}_r + 
	\frac{1}{2})$, its eigenstates are the harmonic-oscillator eigenfunctions
	and form a complete orthonormal basis of $L^2(\mathbb{R}_u)$ (and of any
	unitarily equivalent Hilbert space): on the rotated contour, single-basis
	orthonormalization is automatic.
	
	Analogously, for the outgoing (or decaying-eigenvalue, Gamow--Siegert
	resonant) sector along the conjugate rotated contour $\Gamma_- = e^{+i\pi/4}%
	\mathbb{R}$ parameterized by $x = e^{+i\pi/4}u$ ($u = \sqrt{-i}\,x \in 
	\mathbb{R}$), equation~\eqref{eq:ladder_diff} yields $\widehat{a}_+ = \frac{i%
	}{\sqrt{2}}(u + \frac{d}{du})$ and $\widehat{a}_- = \frac{i}{\sqrt{2}}(u - 
	\frac{d}{du})$. Defining $\widehat{b}_l := -i\widehat{a}_+ = \frac{1}{\sqrt{2%
	}}(u + \frac{d}{du})$ and $\widehat{b}_l^\dagger := -i\widehat{a}_- = \frac{1%
	}{\sqrt{2}}(u - \frac{d}{du})$ preserves $[\widehat{b}_l, \widehat{b}%
	_l^\dagger] = (-i)^2[\widehat{a}_+,\widehat{a}_-] = 1$ with genuine
	adjointness on $L^2(\mathbb{R}_u)$, while $\widehat{a}_+\widehat{a}_- = -%
	\widehat{b}_l\widehat{b}_l^\dagger = -(\widehat{b}_l^\dagger\widehat{b}_l +
	1)$ gives the conjugate skew-adjoint normal Hamiltonian  
	\begin{equation}
		\widehat{H}_- = -i\omega\hbar\left(\widehat{b}_l^\dagger\widehat{b}_l + 
		\frac{1}{2}\right) = -i\omega\hbar\left(-\frac{1}{2}\frac{d^2}{du^2} + \frac{%
			1}{2}u^2\right), \qquad [\widehat{H}_-, \widehat{H}_-^\dagger] = 0.
		\label{eq:H_minus}
	\end{equation}
	
	Because $(\widehat{b}_{r,l}, \widehat{b}_{r,l}^\dagger)$ realize an
	irreducible representation of the Heisenberg--Weyl canonical commutation
	relations on $L^2(\mathbb{R}_u)$, the Stone--von Neumann theorem~\cite%
	{Stone1930,vonNeumann1931,Folland1989,Hall2013} guarantees the existence of
	a unitary equivalence between $L^2(\mathbb{R}_u)$ and the holomorphic
	Segal--Bargmann space $\mathcal{F}(\mathbb{C})$~\cite%
	{Bargmann1961,Bargmann1967,Segal1963,Hall2000,Vourdas2006},  
	\begin{equation}
		\mathcal{F}(\mathbb{C}) = \left\{ \tilde{\psi}(z) \text{ entire on } \mathbb{%
			C} \;\middle|\; \int_{\mathbb{C}} \frac{d^2 z}{\pi}\,e^{-|z|^2}|\tilde{\psi}%
		(z)|^2 < \infty \right\},  \label{eq:bargmann_space}
	\end{equation}
	where $d^2 z = d(\mathrm{Re}\,z)\,d(\mathrm{Im}\,z)$, equipped with the
	Gaussian inner product  
	\begin{equation}
		\langle f | g \rangle_{\mathcal{F}} = \int_{\mathbb{C}} \frac{d^2 z}{\pi}%
		\,e^{-|z|^2}\,\overline{f(z)}\,g(z).  \label{eq:bargmann_inner}
	\end{equation}
	Specifically, the unitary Segal--Bargmann transform $\mathcal{B}: L^2(%
	\mathbb{R}_u) \to \mathcal{F}(\mathbb{C})$, defined by  
	\begin{equation}
		(\mathcal{B}\phi)(z) = \pi^{-1/4}\int_{-\infty}^{\infty}\exp\!\left(-\frac{%
			z^2+u^2}{2} + \sqrt{2}\,zu\right)\phi(u)\,du,  \label{eq:unitary_bargmann}
	\end{equation}
	intertwines the canonical operators according to the Bargmann mapping rules  
	\begin{equation}
		\phi(u) \longmapsto \tilde{\psi}(z), \qquad \widehat{b}_{r,l} \longmapsto 
		\frac{d}{dz}, \qquad \widehat{b}_{r,l}^\dagger \longmapsto z, \qquad 
		\widehat{b}_{r,l}^\dagger\widehat{b}_{r,l} \longmapsto z\frac{d}{dz},
		\label{eq:bargmann_rules}
	\end{equation}
	where $z\frac{d}{dz}$ is the Euler operator. Integration by parts against
	the Gaussian weight $e^{-|z|^2}$ confirms the adjointness relation $\langle
	f | \partial_z g \rangle_{\mathcal{F}} = \langle z f | g \rangle_{\mathcal{F}%
	}$ on $\mathcal{F}(\mathbb{C})$.
	
	Composing $\mathcal{B}$ with the complex coordinate scaling $u = \sqrt{\pm
		i\omega/\hbar}\,\mathrm{x}$ yields a pair of complexified Fresnel--Bargmann
	integral transforms~\cite{Moshinsky1971,Wolf1974} acting directly on
	coordinate-space functions:  
	\begin{equation}
		(\mathcal{B}_{r,l}\psi)(z) = \int_{-\infty}^{\infty} A_{r,l}(z, \mathrm{x}%
		)\,\psi(\mathrm{x})\,d\mathrm{x},  \label{eq:fresnel_bargmann}
	\end{equation}
	with the explicit complex Gaussian kernels  
	\begin{equation}
		\begin{aligned} A_r(z, \mathrm{x}) &=
			\left(\frac{i\omega}{\pi\hbar}\right)^{\frac{1}{4}}\exp\!\left(-%
			\frac{z^2}{2} - \frac{i\omega\mathrm{x}^2}{2\hbar} +
			\sqrt{\frac{2i\omega}{\hbar}}\,z\mathrm{x}\right), \\ A_l(z, \mathrm{x}) &=
			\left(\frac{-i\omega}{\pi\hbar}\right)^{\frac{1}{4}}\exp\!\left(-%
			\frac{z^2}{2} + \frac{i\omega\mathrm{x}^2}{2\hbar} +
			\sqrt{\frac{-2i\omega}{\hbar}}\,z\mathrm{x}\right), \end{aligned}
		\label{eq:bargmann_kernels}
	\end{equation}
	where the real-axis integrals are regularized with $e^{-\epsilon \mathrm{x}%
		^2}$ ($\epsilon\to 0^+$) or evaluated along $\Gamma_\pm = e^{\mp i\pi/4}%
	\mathbb{R}$. Direct integration by parts on equation~%
	\eqref{eq:fresnel_bargmann} confirms the intertwining identities  
	\begin{equation}
		\mathcal{B}_r(\widehat{a}_-\psi)(z) = \frac{d}{dz}(\mathcal{B}_r\psi)(z),
		\qquad \mathcal{B}_r(\widehat{a}_+\psi)(z) = z\,(\mathcal{B}_r\psi)(z),
		\label{eq:intertwine_r}
	\end{equation}
	and, on the conjugate branch,  
	\begin{equation}
		\mathcal{B}_l(-i\widehat{a}_+\psi)(z) = \frac{d}{dz}(\mathcal{B}_l\psi)(z),
		\qquad \mathcal{B}_l(-i\widehat{a}_-\psi)(z) = z\,(\mathcal{B}_l\psi)(z).
		\label{eq:intertwine_l}
	\end{equation}
	In $\mathcal{F}(\mathbb{C})$, the IHO Hamiltonian therefore takes the two
	skew-adjoint normal representations  
	\begin{equation}
		\widehat{H}_{\mathcal{F}}^{(r)} = +i\omega\hbar\left(z\frac{d}{dz} + \frac{1%
		}{2}\right), \qquad \widehat{H}_{\mathcal{F}}^{(l)} = -i\omega\hbar\left(z%
		\frac{d}{dz} + \frac{1}{2}\right).  \label{eq:H_bargmann}
	\end{equation}
	
	\subsection{Holomorphic eigenvalue equation, quantization, and orthonormal
		Fock basis}
	
	Under the representations~\eqref{eq:H_bargmann}, the stationary eigenvalue
	equation $\widehat{H}_{\mathcal{F}}^{(\pm)}\tilde{\psi}(z) = E\tilde{\psi}(z)
	$ (identifying $\widehat{H}_{\mathcal{F}}^{(+)} \equiv \widehat{H}_{\mathcal{%
			F}}^{(r)}$ and $\widehat{H}_{\mathcal{F}}^{(-)} \equiv \widehat{H}_{\mathcal{%
			F}}^{(l)}$) reduces to the separable first-order ordinary differential
	equation  
	\begin{equation}
		\pm i\omega\hbar\left(z\frac{d}{dz} + \frac{1}{2}\right)\tilde{\psi}(z) = E%
		\tilde{\psi}(z) \quad \Longrightarrow \quad z\frac{d\tilde{\psi}}{dz} =
		\left(\mp i\frac{E}{\omega\hbar} - \frac{1}{2}\right)\tilde{\psi}(z).
		\label{eq:bargmann_ode}
	\end{equation}
	Its general solution is the power function  
	\begin{equation}
		\tilde{\psi}(z) = C z^\lambda, \qquad \lambda = \mp i\frac{E}{\omega\hbar} - 
		\frac{1}{2}.  \label{eq:bargmann_general_sol}
	\end{equation}
	
	In the Bargmann representation, quantization follows immediately from a
	purely analytic requirement: every physical vector $\tilde{\psi}(z) \in 
	\mathcal{F}(\mathbb{C})$ must be an \emph{entire function} on $\mathbb{C}$~%
	\cite{Bargmann1961}. In order to avoid a branch point or pole singularity at 
	$z=0$, the exponent $\lambda$ in equation~\eqref{eq:bargmann_general_sol}
	must be a non-negative integer:  
	\begin{equation}
		\lambda = n, \qquad n = 0,1,2,\dots.  \label{eq:quant_condition}
	\end{equation}
	Substituting $\lambda = n$ back into equation~\eqref{eq:bargmann_general_sol}
	reproduces the exact discrete conjugate pure imaginary spectrum of the IHO:  
	\begin{equation}
		E_n^\pm = \pm i\omega\hbar\left(n + \frac{1}{2}\right).
		\label{eq:bargmann_spectrum}
	\end{equation}
	
	Because $\widehat{H}_{\mathcal{F}}^{(\pm)}$ are skew-adjoint normal
	operators on $\mathcal{F}(\mathbb{C})$, the resulting monomial
	eigenfunctions $\tilde{\psi}_n(z) = C_n z^n$ are directly normalizable and
	mutually orthogonal in the single Bargmann inner product~%
	\eqref{eq:bargmann_inner}, without introducing any biorthogonal dual basis
	inside $\mathcal{F}(\mathbb{C})$. Passing to polar coordinates $z =
	re^{i\theta}$ with $d^2 z = r\,dr\,d\theta$, we evaluate  
	\begin{equation}
		\langle \tilde{\psi}_m | \tilde{\psi}_n \rangle_{\mathcal{F}} = \frac{C_m^*
			C_n}{\pi}\int_{\mathbb{C}} d^2 z\,e^{-|z|^2}(\bar{z})^m z^n = \frac{C_m^* C_n%
		}{\pi}\int_0^\infty r\,dr\,e^{-r^2}r^{m+n}\int_0^{2\pi}
		d\theta\,e^{i(n-m)\theta}.  \label{eq:bargmann_ortho_calc}
	\end{equation}
	The angular integral yields $2\pi\delta_{mn}$, while for $m=n$ the radial
	integral gives $\int_0^\infty 2r\,e^{-r^2}r^{2n}\,dr = \Gamma(n+1) = n!$.
	Enforcing $\langle \tilde{\psi}_n | \tilde{\psi}_n \rangle_{\mathcal{F}} =
	|C_n|^2 n! = 1$ yields $C_n = 1/\sqrt{n!}$, so that the normalized
	eigenstates in $\mathcal{F}(\mathbb{C})$ form the canonical Fock monomial
	basis  
	\begin{equation}
		\tilde{\psi}_n(z) = \frac{z^n}{\sqrt{n!}}, \qquad \langle \tilde{\psi}_m | 
		\tilde{\psi}_n \rangle_{\mathcal{F}} = \delta_{mn}.  \label{eq:bargmann_fock}
	\end{equation}
	Physically, the Gaussian weight $e^{-|z|^2}$ in equation~%
	\eqref{eq:bargmann_inner} suppresses the algebraic growth of $z^n$ as $%
	|z|\to\infty$, playing in phase space the exact regularizing role that
	complex scaling and the Fresnel measure play for non-$L^2(\mathbb{R})$
	Siegert resonance tails~\cite{Siegert1939,Civitarese2004,Moiseyev2011} in
	coordinate space.
	
	\subsection{Inverse Bargmann transform and recovery of complex Hermite wave
		functions}
	
	We can invert the Bargmann eigenstates~\eqref{eq:bargmann_fock} back to the
	coordinate representation in two equivalent ways, confirming complete
	consistency with equations~\eqref{eq:32}--\eqref{eq:33}.
	
	First, algebraically, under the intertwining relations~%
	\eqref{eq:intertwine_r}, the Bargmann vacuum $\tilde{\psi}_0(z) = 1$
	satisfying $\frac{d}{dz}\tilde{\psi}_0(z) = 0$ maps in coordinate space to $%
	\widehat{a}_-\psi_0^r(\mathrm{x}) = 0$, i.e., $(\mathrm{x} - i\frac{\hbar}{%
		\omega}\frac{d}{d\mathrm{x}})\psi_0^r(\mathrm{x}) = 0$, giving $\psi_0^r(%
	\mathrm{x}) = (i\omega/(\pi\hbar))^{1/4} e^{-i\omega\mathrm{x}^2/(2\hbar)}$.
	Likewise, the Fock monomial generation $\tilde{\psi}_n(z) = \frac{z^n}{\sqrt{%
			n!}}\tilde{\psi}_0(z)$ maps back via $z \leftrightarrow \widehat{a}_+ = 
	\sqrt{\frac{i}{2}}\,(\sqrt{\omega/\hbar}\,\mathrm{x} + i\sqrt{\hbar/\omega}%
	\,d/d\mathrm{x})$ directly to  
	\begin{equation}
		\psi_n^r(\mathrm{x}) = \frac{(\widehat{a}_+)^n}{\sqrt{n!}}\psi_0^r(\mathrm{x}%
		) = \frac{1}{\sqrt{n!}}\left(\sqrt{\frac{i}{2}}\right)^n \left(\frac{i\omega%
		}{\pi\hbar}\right)^{\frac{1}{4}}\left(\sqrt{\frac{\omega}{\hbar}}\,\mathrm{x}
		+ i\sqrt{\frac{\hbar}{\omega}}\frac{d}{d\mathrm{x}}\right)^n e^{-i\frac{%
				\omega\mathrm{x}^2}{2\hbar}},  \label{eq:inverse_alg}
	\end{equation}
	which coincides identically with equation~\eqref{eq:32} (and similarly under 
	$\mathcal{B}_l$, $z \leftrightarrow -i\widehat{a}_-$ yields equation~%
	\eqref{eq:33}). Here the ground state $\psi_0^r(\mathrm{x}) \sim e^{-i\omega%
		\mathrm{x}^2/(2\hbar)}$ has probability current $j_r(\mathrm{x}) = \hbar\,%
	\mathrm{Im}([\psi_0^r]^*\partial_{\mathrm{x}}\psi_0^r) = -\omega\mathrm{x}%
	|\psi_0^r|^2$ and thus represents an incoming anti-resonant wave flowing
	toward the unstable potential top (growing as $e^{+\omega t/2}$ in forward
	time), whereas its conjugate partner $\psi_0^l(\mathrm{x}) = [\psi_0^r(%
	\mathrm{x})]^* \sim e^{i\omega\mathrm{x}^2/(2\hbar)}$ carries $j_l(\mathrm{x}%
	) = +\omega\mathrm{x}|\psi_0^l|^2$ and represents the outgoing
	Gamow--Siegert wave carrying probability away from the potential top
	(decaying as $e^{-\omega t/2}$ in forward time), in full consistency with
	the Bargmann semigroup domains in equation~\eqref{eq:bargmann_prop} and the
	spectral branch splitting in section~\ref{sec:propagator}.
	
	Second, analytically, the exact inverse Fresnel--Bargmann transforms from $%
	\mathcal{F}(\mathbb{C})$ back to coordinate space are given by  
	\begin{equation}
		\psi_n^{r,l}(\mathrm{x}) = (\mathcal{B}_{r,l}^{-1}\tilde{\psi}_n)(\mathrm{x}%
		) = \int_{\mathbb{C}} \frac{d^2 z}{\pi}\,e^{-|z|^2}\,A_{r,l}(\bar{z}, 
		\mathrm{x})\,\tilde{\psi}_n(z).  \label{eq:inverse_bargmann}
	\end{equation}
	To evaluate equation~\eqref{eq:inverse_bargmann}, we recall the Hermite
	generating function  
	\begin{equation}
		\exp(2wt - t^2) = \sum_{m=0}^{\infty}\mathcal{H}_m^{\mathrm{std}}(w)\,\frac{%
			t^m}{m!},
	\end{equation}
	and set $w = \sqrt{\frac{\pm i\omega}{\hbar}}\,\mathrm{x}$ and $t = \bar{z}/%
	\sqrt{2}$. Using equation~\eqref{eq:complex_H}, the kernels~%
	\eqref{eq:bargmann_kernels} expand directly in terms of the dual coordinate
	eigenstates~\eqref{eq:32}--\eqref{eq:33}:  
	\begin{equation}
		A_{r,l}(\bar{z}, \mathrm{x}) = \sum_{m=0}^{\infty}\psi_m^{r,l}(\mathrm{x})\,%
		\overline{\tilde{\psi}_m(z)}.  \label{eq:kernel_expansion}
	\end{equation}
	Inserting equation~\eqref{eq:kernel_expansion} into equation~%
	\eqref{eq:inverse_bargmann} and invoking the Bargmann orthogonality~%
	\eqref{eq:bargmann_fock} immediately yields  
	\begin{equation}
		(\mathcal{B}_{r,l}^{-1}\tilde{\psi}_n)(\mathrm{x}) = \sum_{m=0}^{\infty}
		\psi_m^{r,l}(\mathrm{x})\,\langle\tilde{\psi}_m|\tilde{\psi}_n\rangle_{%
			\mathcal{F}} = \psi_n^{r,l}(\mathrm{x}),  \label{eq:inverse_exact}
	\end{equation}
	recovering both the complex Hermite polynomial factors and the exact
	normalization constants of equations~\eqref{eq:32}--\eqref{eq:33}.
	Furthermore, on the rotated contour $\Gamma_+ = e^{-i\pi/4}\mathbb{R}$ with $%
	\mathrm{x} = e^{-i\pi/4}\sqrt{\hbar/\omega}\,u$, the rescaled functions $%
	\phi_m(u) = (i\omega/\hbar)^{-1/4}\psi_m^r(\mathrm{x})$ reduce to the real
	harmonic-oscillator eigenfunctions $(2^m m!\sqrt{\pi})^{-1/2}e^{-u^2/2}%
	\mathcal{H}_m^{\mathrm{std}}(u)$ for $u\in\mathbb{R}$, while $du = \sqrt{%
		i\omega/\hbar}\,d\mathrm{x}$. Consequently, the unitary Bargmann inner
	product equals the $L^2(\mathbb{R}_u)$ inner product on $\Gamma_+$,  
	\begin{equation}
		\langle \tilde{\psi}_m | \tilde{\psi}_n \rangle_{\mathcal{F}} =
		\int_{-\infty}^{\infty} \overline{\phi_m(u)}\,\phi_n(u)\,du =
		\int_{-\infty}^{\infty} \phi_m(u)\,\phi_n(u)\,du = \int_{\Gamma_+} \psi_m^r(%
		\mathrm{x})\,\psi_n^r(\mathrm{x})\,d\mathrm{x},
	\end{equation}
	which upon deforming $\Gamma_+$ back to the real axis $\mathbb{R}$ (with
	Fresnel convergence factor $e^{-\epsilon \mathrm{x}^2}$, $\epsilon\to 0^+$)
	and using $\psi_m^r(\mathrm{x}) = [\psi_m^l(\mathrm{x})]^*$ for $\mathrm{x}%
	\in\mathbb{R}$ reduces precisely to the biorthogonal pairing $%
	\int_{-\infty}^{\infty}[\psi_m^l(\mathrm{x})]^*\psi_n^r(\mathrm{x})\,d%
	\mathrm{x} = \delta_{mn}$ (the $c$-product of non-Hermitian resonance theory~%
	\cite{Moiseyev2011}).
	
	\subsection{Evolution semigroups and reproducing-kernel coherent states in 
		\texorpdfstring{$\mathcal{F}(\mathbb{C})$}{F(C)}}
	
	The Bargmann representation also provides transparent insight into the
	time-evolution and coherent-state structures analyzed in the following
	sections. Since $z\frac{d}{dz}$ generates complex dilatations, $e^{\tau z%
		\frac{d}{dz}}f(z) = f(e^\tau z)$, the time-evolution operators generated by $%
	\widehat{H}_{\mathcal{F}}^{(l)}$ and $\widehat{H}_{\mathcal{F}}^{(r)}$ act
	on $f\in\mathcal{F}(\mathbb{C})$ simply as weighted rescalings:  
	\begin{equation}
		e^{-\frac{i}{\hbar}\widehat{H}_{\mathcal{F}}^{(l)}T}f(z) = e^{-\frac{1}{2}%
			\omega T}f(e^{-\omega T}z), \qquad e^{-\frac{i}{\hbar}\widehat{H}_{\mathcal{F%
			}}^{(r)}T}f(z) = e^{+\frac{1}{2}\omega T}f(e^{+\omega T}z).
		\label{eq:bargmann_semigroup}
	\end{equation}
	Because $\|e^{-\frac{i}{\hbar}\widehat{H}_{\mathcal{F}}^{(l)}T}\tilde{\psi}%
	_n\|_{\mathcal{F}} = e^{-\omega(n+1/2)T}$, the decaying branch $\widehat{H}_{%
		\mathcal{F}}^{(l)}$ generates a bounded contraction semigroup on $\mathcal{F}%
	(\mathbb{C})$ exclusively for forward times $T \ge 0$, whereas the growing
	branch $\widehat{H}_{\mathcal{F}}^{(r)}$ generates a bounded contraction
	semigroup exclusively for backward times $T \le 0$. Using the reproducing
	kernel $e^{z\bar{w}}$ of $\mathcal{F}(\mathbb{C})$~\cite{Bargmann1961}, the
	corresponding evolution kernels in Bargmann space are elementary
	exponentials:  
	\begin{equation}
		\begin{aligned} K_{\mathcal{F}}^{(l)}(z,\bar{w};T) &= e^{-\frac{1}{2}\omega
				T}\exp\!\left(e^{-\omega T}z\bar{w}\right) \qquad (T>0), \\
			K_{\mathcal{F}}^{(r)}(z,\bar{w};T) &= e^{+\frac{1}{2}\omega
				T}\exp\!\left(e^{+\omega T}z\bar{w}\right) \qquad (T<0), \end{aligned}
		\label{eq:bargmann_prop}
	\end{equation}
	whose inverse Bargmann transforms via equation~\eqref{eq:kernel_expansion}
	lead directly to the dual spectral expansions of the Feynman propagator
	established in section~\ref{sec:propagator}. Likewise, coherent states
	solving $\widehat{a}_-|\alpha\rangle_r = \alpha|\alpha\rangle_r$ map under $%
	\mathcal{B}_r$ to the entire exponential functions $\tilde{\psi}_\alpha(z)
	\propto e^{\alpha z}$ in $\mathcal{F}(\mathbb{C})$, which are immediate
	eigenfunctions of $\frac{d}{dz}$ and serve as the starting point for the
	bi-coherent states of section~\ref{sec:coherent}.
	
	\section{Feynman propagator and time-asymmetric resonance reconstruction}
	
	\label{sec:propagator}
	
	We now turn to the central result of this work. Although the real-time
	Feynman propagator is unique and conventionally associated with the
	continuous real spectrum of $\widehat{H}$ on $L^2(\mathbb{R})$, the discrete
	resonance states~\eqref{eq:32}--\eqref{eq:33} of sections~\ref{sec:prelim}--%
	\ref{sec:bargmann} come in two disjoint branches and carry no trace of the
	real scattering continuum. We show below that each discrete branch
	nevertheless rebuilds the full unitary kernel by itself in its respective
	time domain, in exact correspondence with the Bargmann-space semigroups~%
	\eqref{eq:bargmann_prop}. To provide an independent, self-contained
	benchmark for this spectral reconstruction, we begin with a direct
	path-integral evaluation of the real-time propagator.
	
	\subsection{Feynman path integral and the propagator of the inverted
		harmonic oscillator}
	
	In Feynman path-integral theory~\cite{Liang2023,Feynman1965}, the real-time
	propagator of the IHO is defined by  
	\begin{equation}
		K(\mathrm{x}_T, T; \mathrm{x}_0, 0) = \int \mathcal{D}\{\mathrm{x}\}\, e^{%
			\frac{i}{\hbar}\int_0^T \left(\frac{1}{2}\dot{\mathrm{x}}^2 - V(\mathrm{x}%
			)\right)dt}, \qquad V(\mathrm{x}) = -\frac{1}{2}\omega^2\mathrm{x}^2,
		\label{eq:34}
	\end{equation}
	where $\mathcal{D}\{\mathrm{x}\} = \lim_{N\to\infty}\sqrt{\frac{1}{2\pi
			i\hbar\epsilon_N}}\prod_{n=1}^{N-1}\left(\sqrt{\frac{1}{2\pi i\hbar\epsilon_n%
	}}\,d\mathrm{x}_n\right)$ with time slices $\epsilon_n = t_n - t_{n-1}$, $%
	t_0 = 0$, and $t_N = T$. For a free particle ($V(\mathrm{x}) = 0$), equation~%
	\eqref{eq:34} reduces to  
	\begin{equation}
		K_0(\mathrm{x}_T, T; \mathrm{x}_0, 0) = \sqrt{\frac{1}{2\pi i\hbar T}}\,e^{%
			\frac{i}{2T\hbar}(\mathrm{x}_T - \mathrm{x}_0)^2}.  \label{eq:35}
	\end{equation}
	
	Rewriting the inverted-oscillator action in equation~\eqref{eq:34} by
	completing the square,  
	\begin{equation}
		K(\mathrm{x}_T, T; \mathrm{x}_0, 0) = \int \mathcal{D}\{\mathrm{x}\}\,
		e^{\int_0^T \frac{i}{2\hbar}\left[(\dot{\mathrm{x}} - \omega\mathrm{x})^2 +
			\omega\frac{d\mathrm{x}^2}{dt}\right]dt} = e^{\frac{i\omega}{2\hbar}(\mathrm{%
				x}_T^2 - \mathrm{x}_0^2)} \int \mathcal{D}\{\mathrm{x}\}\, e^{\int_0^T \frac{%
				i}{2\hbar}(\dot{\mathrm{x}} - \omega\mathrm{x})^2 dt},
	\end{equation}
	we adapt Moriconi's Euclidean transformation~\cite{Moriconi2004} to real
	time via the coordinate substitution $\mathrm{x}(t) = z(t)e^{\omega t}$
	(here $z(t)\in\mathbb{R}$ denotes the transformed real path variable,
	distinct from the complex Bargmann coordinate in section~\ref{sec:bargmann}%
	), under which  
	\begin{equation}
		K(\mathrm{x}_T, T; \mathrm{x}_0, 0) = e^{\frac{i\omega}{2\hbar}(\mathrm{x}%
			_T^2 - \mathrm{x}_0^2)} \int \mathcal{D}\{z e^{\omega t}\}\, e^{\int_0^T 
			\frac{i}{2\hbar}\dot{z}^2 e^{2\omega t}dt}.  \label{eq:36}
	\end{equation}
	Introducing the rescaled time variable  
	\begin{equation}
		\tau = -\frac{e^{-2\omega t}}{2\omega}, \qquad d\tau = e^{-2\omega t}dt,
		\qquad \frac{dz(t)}{dt} = \frac{dz(\tau)}{d\tau}e^{-2\omega t},
	\end{equation}
	converts the action into free-particle form in $\tau$:  
	\begin{equation}
		\int_0^T \frac{i}{2\hbar}\dot{z}^2 e^{2\omega t}dt = \int_{\tau_0}^{\tau_T} 
		\frac{i}{2\hbar}\left(\frac{dz(\tau)}{d\tau}\right)^2 d\tau.
	\end{equation}
	Since the time-slice intervals transform as $\epsilon_n(t) = e^{2\omega\bar{t%
		}_n}\epsilon_n(\tau)$ with midpoint $\bar{t}_n = (t_n + t_{n-1})/2$ as $%
	N\to\infty$, the path-integral measure in equation~\eqref{eq:36} acquires
	the Jacobian factor  
	\begin{equation}
		\mathcal{D}\{z(t)e^{\omega t}\} = \lim_{N\to\infty}\left(\prod_{n=1}^{N-1}e^{%
			\frac{1}{2}\omega\epsilon_n(t)}\right)e^{-\omega\bar{t}_N}\mathcal{D}%
		\{z(\tau)\} = e^{-\frac{\omega T}{2}}\mathcal{D}\{z(\tau)\}.
	\end{equation}
	Consequently, the propagator reduces to that of a free particle in $%
	(z(\tau), \tau)$:  
	\begin{equation}
		\begin{aligned} K(\mathrm{x}_T, T; \mathrm{x}_0, 0) &=
			e^{\frac{i\omega}{2\hbar}(\mathrm{x}_T^2 - \mathrm{x}_0^2) - \frac{\omega
					T}{2}} \int \mathcal{D}\{z(\tau)\}\,
			e^{\int_{\tau_0}^{\tau_T}\frac{i}{2\hbar}\dot{z}^2(\tau)d\tau} \\ &=
			e^{\frac{i\omega}{2\hbar}(\mathrm{x}_T^2 - \mathrm{x}_0^2) - \frac{\omega
					T}{2}} \sqrt{\frac{1}{2\pi i\hbar(\tau_T - \tau_0)}}\,
			e^{\frac{i}{2\hbar(\tau_T - \tau_0)}[z(\tau_T) - z(\tau_0)]^2}. \end{aligned}
	\end{equation}
	Substituting $\tau_T - \tau_0 = \frac{1 - e^{-2\omega T}}{2\omega} = \frac{%
		e^{-\omega T}\sinh\omega T}{\omega}$, $z(\tau_T) = \mathrm{x}_T e^{-\omega T}
	$, and $z(\tau_0) = \mathrm{x}_0$ recovers the exact inverted-oscillator
	propagator~\cite{Barton1986}  
	\begin{equation}
		K(\mathrm{x}_T, T; \mathrm{x}_0, 0) = \sqrt{\frac{\omega}{2\pi
				i\hbar\sinh\omega T}}\, e^{\frac{i\omega}{2\hbar\sinh(\omega T)}\left[(%
			\mathrm{x}_T^2 + \mathrm{x}_0^2)\cosh(\omega T) - 2\mathrm{x}_T\mathrm{x}_0%
			\right]}.  \label{eq:37}
	\end{equation}
	
	\subsection{Dual resonance expansions and Mehler summation}
	
	Having obtained the closed-form benchmark~\eqref{eq:37} directly from the
	path integral, we now demonstrate how it is reconstructed exclusively from
	the discrete imaginary-eigenvalue ladder. From the operator definition of
	the propagator,  
	\begin{equation}
		K(\mathrm{x}_T, T; \mathrm{x}_0, 0) = \langle \mathrm{x}_T | e^{-\frac{i}{%
				\hbar}\widehat{H}T} | \mathrm{x}_0 \rangle,
	\end{equation}
	inserting the two dual completeness relations $\sum_{n=0}^{\infty}|n%
	\rangle_l\,{}_r\langle n| = 1$ and $\sum_{n=0}^{\infty}|n\rangle_r\,{}_l%
	\langle n| = 1$ (valid on the respective rigged Hilbert space test domains~%
	\cite{Chruscinski2004}, and equivalently induced by inverting the forward
	and backward Bargmann semigroup kernels~\eqref{eq:bargmann_prop} via the
	kernel expansion~\eqref{eq:kernel_expansion}) together with equation~%
	\eqref{eq:15} yields two dual spectral expansions:  
	\begin{equation}
		K(\mathrm{x}_T, T; \mathrm{x}_0, 0) = \sum_{n=0}^{\infty} e^{-\omega\left(n
			+ \frac{1}{2}\right)T} \psi_n^l(\mathrm{x}_T) \left(\psi_n^r(\mathrm{x}%
		_0)\right)^*,  \label{eq:38}
	\end{equation}
	and  
	\begin{equation}
		K(\mathrm{x}_T, T; \mathrm{x}_0, 0) = \sum_{n=0}^{\infty} e^{\omega\left(n + 
			\frac{1}{2}\right)T} \psi_n^r(\mathrm{x}_T) \left(\psi_n^l(\mathrm{x}%
		_0)\right)^*.  \label{eq:39}
	\end{equation}
	
	For forward propagation ($T > 0$), substituting equations~\eqref{eq:32}--%
	\eqref{eq:33} into the convergent expansion~\eqref{eq:38} gives  
	\begin{equation}
		K(\mathrm{x}_T, T; \mathrm{x}_0, 0) = \sqrt{\frac{-i\omega}{\pi\hbar}}\, e^{i%
			\frac{\omega(\mathrm{x}_T^2 + \mathrm{x}_0^2)}{2\hbar}} e^{-\frac{\omega}{2}%
			T} \sum_{n=0}^{\infty} \frac{(-i e^{-\omega T})^n}{n! 2^n} H_n^*\!\left(%
		\sqrt{\frac{\omega}{\hbar}}\,\mathrm{x}_T\right) H_n^*\!\left(\sqrt{\frac{%
				\omega}{\hbar}}\,\mathrm{x}_0\right).
	\end{equation}
	Using Mehler's formula~\cite{Lebedev1963,Liang1991} and equation~%
	\eqref{eq:complex_H}, the bilinear generating function of the complex
	Hermite polynomials for $|f(\omega)| < 1$ is  
	\begin{equation}
		\begin{aligned} G(\mathrm{x}_T, \mathrm{x}_0, f(\omega)) &=
			\sum_{n=0}^{\infty} \frac{[if(\omega)]^n}{n! 2^n}
			H_n\!\left(\sqrt{\frac{\omega}{\hbar}}\,\mathrm{x}_T\right)
			H_n\!\left(\sqrt{\frac{\omega}{\hbar}}\,\mathrm{x}_0\right) \\ &=
			\frac{1}{\sqrt{1 - f^2(\omega)}}
			\exp\left\{\frac{i\omega\left[2f(\omega)\mathrm{x}_T\mathrm{x}_0 -
				f^2(\omega)(\mathrm{x}_T^2 + \mathrm{x}_0^2)\right]}{\hbar[1 -
				f^2(\omega)]}\right\}, \end{aligned}  \label{eq:G_mehler}
	\end{equation}
	whose complex conjugate for real $f(\omega)$ (conjugating both $i \to -i$
	and $H_n \to H_n^*$) is  
	\begin{equation}
		G^*(\mathrm{x}_T, \mathrm{x}_0, f(\omega)) = \sum_{n=0}^{\infty} \frac{%
			[-if(\omega)]^n}{n! 2^n} H_n^*\!\left(\sqrt{\frac{\omega}{\hbar}}\,\mathrm{x}%
		_T\right) H_n^*\!\left(\sqrt{\frac{\omega}{\hbar}}\,\mathrm{x}_0\right).
	\end{equation}
	Setting $f(\omega) = e^{-\omega T} \in (0,1)$, equation~\eqref{eq:38} sums
	in closed form via $G^*$ to  
	\begin{equation}
		\begin{aligned} K(\mathrm{x}_T, T; \mathrm{x}_0, 0) &=
			\sqrt{\frac{-i\omega}{\pi\hbar}}\, e^{i\frac{\omega(\mathrm{x}_T^2 +
					\mathrm{x}_0^2)}{2\hbar}} e^{-\frac{\omega}{2}T} G^*\!\left(\mathrm{x}_T,
			\mathrm{x}_0, e^{-\omega T}\right) \\ &= \sqrt{\frac{\omega}{2\pi
					i\hbar\sinh\omega T}}\, e^{\frac{i\omega\left[(\mathrm{x}_T^2 +
					\mathrm{x}_0^2)\cosh\omega T -
					2\mathrm{x}_T\mathrm{x}_0\right]}{2\hbar\sinh\omega T}}, \end{aligned}
	\end{equation}
	which coincides identically with equation~\eqref{eq:37}.
	
	Conversely, the second expansion~\eqref{eq:39} reads  
	\begin{equation}
		K(\mathrm{x}_T, T; \mathrm{x}_0, 0) = \sqrt{\frac{i\omega}{\pi\hbar}}\, e^{-i%
			\frac{\omega(\mathrm{x}_T^2 + \mathrm{x}_0^2)}{2\hbar}} e^{\frac{1}{2}\omega
			T} \sum_{n=0}^{\infty} \frac{(i e^{\omega T})^n}{2^n n!} H_n\!\left(\sqrt{%
			\frac{\omega}{\hbar}}\,\mathrm{x}_T\right) H_n\!\left(\sqrt{\frac{\omega}{%
				\hbar}}\,\mathrm{x}_0\right).
	\end{equation}
	For backward propagation $T < 0$ ($f(\omega) = e^{\omega T} \in (0,1)$),
	applying $G(\mathrm{x}_T, \mathrm{x}_0, e^{\omega T})$ from equation~%
	\eqref{eq:G_mehler} yields  
	\begin{equation}
		\begin{aligned} K(\mathrm{x}_T, T; \mathrm{x}_0, 0) &=
			\sqrt{\frac{i\omega}{\pi\hbar}}\, e^{-i\frac{\omega(\mathrm{x}_T^2 +
					\mathrm{x}_0^2)}{2\hbar}} e^{\frac{1}{2}\omega T} G\!\left(\mathrm{x}_T,
			\mathrm{x}_0, e^{\omega T}\right) \\ &= \sqrt{\frac{\omega}{2\pi
					i\hbar\sinh\omega T}}\, e^{\frac{i\omega\left[(\mathrm{x}_T^2 +
					\mathrm{x}_0^2)\cosh\omega T -
					2\mathrm{x}_T\mathrm{x}_0\right]}{2\hbar\sinh\omega T}}, \end{aligned}
	\end{equation}
	reproducing the exact same propagator~\eqref{eq:37}. Notably, equation~%
	\eqref{eq:38} converges exclusively for forward time evolution ($T>0$),
	whereas equation~\eqref{eq:39} converges exclusively for backward time
	evolution ($T<0$) with the square-root branch $\sqrt{i\omega/(-2\pi\hbar%
		\sinh\omega T)} = e^{i\pi/4}\sqrt{\omega/(2\pi\hbar|\sinh\omega T|)}$
	matching the $T\to 0^-$ Fresnel normalization of the backward propagator.
	This forward\slash backward branch splitting reflects both the
	time-asymmetric semigroup structure of decaying resonant and growing
	anti-resonant states in rigged Hilbert space~\cite%
	{Shimbori2000a,Chruscinski2004} and the contraction semigroup domains of the
	Bargmann kernels~\eqref{eq:bargmann_prop}, confirming full consistency
	between the path-integral and spectral formulations. The unitary real-time
	kernel of a Hamiltonian whose $L^2(\mathbb{R})$ spectrum is purely
	continuous and real is thus rebuilt, term by term, from the discrete
	imaginary ladder alone, with no contribution from the scattering continuum.
	
	\section{Coherent states of imaginary-frequency boson operators}
	
	\label{sec:coherent}
	
	We now construct the coherent states of the imaginary-frequency boson
	operators as eigenstates of the respective lowering operators $\widehat{a}_-$
	and $\widehat{a}_+$ for the ``ket'' and ``bra'' sets, providing an explicit,
	analytically tractable realization of bi-coherent states for $\mathcal{D}$%
	-pseudo-bosons~\cite{Bagarello2022} with the elementary holomorphic image $%
	e^{\alpha z}$ in $\mathcal{F}(\mathbb{C})$. Coherent states of the IHO in a
	pseudo-Hermitian metric were constructed in~\cite%
	{Zerimeche2023,Amaouche2022b}.
	
	\subsection{Coherent states of the ``ket'' and ``bra'' sets}
	
	The ``ket'' coherent state $|\alpha\rangle_r$ is defined as the eigenstate
	of the lowering operator $\widehat{a}_-$:  
	\begin{equation}
		\widehat{a}_-|\alpha\rangle_r = \alpha|\alpha\rangle_r.  \label{eq:40}
	\end{equation}
	Expanding $|\alpha\rangle_r = \sum_{n=0}^{\infty} c_n^r |n\rangle_r$ and
	using $\widehat{a}_-|n\rangle_r = \sqrt{n}|n-1\rangle_r$ from equation~%
	\eqref{eq:19}, equation~\eqref{eq:40} gives  
	\begin{equation}
		\sum_{n=1}^{\infty} c_n^r \sqrt{n}|n-1\rangle_r = \alpha\sum_{n=0}^{\infty}
		c_n^r |n\rangle_r,  \label{eq:41}
	\end{equation}
	which yields the recurrence relation  
	\begin{equation}
		c_n^r = \frac{\alpha}{\sqrt{n}}\,c_{n-1}^r = \frac{\alpha^n}{\sqrt{n!}}%
		\,c_0^r, \qquad |\alpha\rangle_r = c_0^r \sum_{n=0}^{\infty} \frac{\alpha^n}{%
			\sqrt{n!}}|n\rangle_r.  \label{eq:45}
	\end{equation}
	Under the Bargmann mapping $\mathcal{B}_r$ of section~\ref{sec:bargmann}, $%
	|n\rangle_r \mapsto z^n/\sqrt{n!}$, so $|\alpha\rangle_r$ corresponds
	directly to the entire exponential $\tilde{\psi}_\alpha^r(z) = c_0^r
	e^{\alpha z} \in \mathcal{F}(\mathbb{C})$, for which equation~\eqref{eq:40}
	reduces immediately to $\frac{d}{dz}e^{\alpha z} = \alpha e^{\alpha z}$.
	
	Similarly, the ``bra'' coherent state $|\alpha\rangle_l$ is defined as the
	eigenstate of its lowering operator $\widehat{a}_+$:  
	\begin{equation}
		\widehat{a}_+|\alpha\rangle_l = \alpha|\alpha\rangle_l, \qquad
		|\alpha\rangle_l = \sum_{n=0}^{\infty} c_n^l |n\rangle_l.  \label{eq:46}
	\end{equation}
	Using $\widehat{a}_+|n\rangle_l = i\sqrt{n}|n-1\rangle_l$ from equation~%
	\eqref{eq:19}, we obtain  
	\begin{equation}
		\sum_{n=1}^{\infty} c_n^l i\sqrt{n}|n-1\rangle_l = \alpha
		\sum_{n=0}^{\infty} c_n^l |n\rangle_l \quad \Longrightarrow \quad c_n^l = 
		\frac{-i\alpha}{\sqrt{n}}\,c_{n-1}^l = \frac{(-i\alpha)^n}{\sqrt{n!}}\,c_0^l.
		\label{eq:49}
	\end{equation}
	Imposing the biorthogonal normalization condition between the dual coherent
	states,  
	\begin{equation}
		{}_l\langle \alpha|\alpha\rangle_r = (c_0^l)^* c_0^r \sum_{n=0}^{\infty} 
		\frac{i^n|\alpha|^{2n}}{n!} = (c_0^l)^* c_0^r e^{i|\alpha|^2} = 1,
	\end{equation}
	determines the symmetric normalization factors $c_0^r = e^{-\frac{i}{2}%
		|\alpha|^2}$ and $c_0^l = e^{\frac{i}{2}|\alpha|^2}$. The normalized dual
	coherent states thus read  
	\begin{equation}
		|\alpha\rangle_r = e^{-\frac{i}{2}|\alpha|^2} \sum_{n=0}^{\infty} \frac{%
			\alpha^n}{\sqrt{n!}}|n\rangle_r, \qquad |\alpha\rangle_l = e^{\frac{i}{2}%
			|\alpha|^2} \sum_{n=0}^{\infty} \frac{(-i\alpha)^n}{\sqrt{n!}}|n\rangle_l.
		\label{eq:51}
	\end{equation}
	
	\subsection{Minimum uncertainty}
	
	To evaluate the uncertainty relation in the dual coherent states, we invert
	equation~\eqref{eq:9} to express the dimensionless position and momentum
	operators as  
	\begin{equation}
		\widehat{x} = \frac{1}{\sqrt{2i}}(\widehat{a}_- + \widehat{a}_+), \qquad 
		\widehat{p} = \frac{1}{\sqrt{2i}}(\widehat{a}_- - \widehat{a}_+).
		\label{eq:53}
	\end{equation}
	Using equations~\eqref{eq:40} and \eqref{eq:46} together with $\widehat{a}%
	_\pm^\dagger = -i\widehat{a}_\pm$ from equation~\eqref{eq:10}, the right and
	left eigenvalue relations are  
	\begin{equation}
		\begin{aligned} \widehat{a}_-|\alpha\rangle_r &= \alpha|\alpha\rangle_r, &
			{}_r\langle\alpha|\widehat{a}_-^\dagger &= {}_r\langle\alpha|\alpha^*
			\;\Rightarrow\; {}_r\langle\alpha|\widehat{a}_- =
			{}_r\langle\alpha|(i\alpha^*), \\ \widehat{a}_+|\alpha\rangle_l &=
			\alpha|\alpha\rangle_l, & {}_l\langle\alpha|\widehat{a}_+^\dagger &=
			{}_l\langle\alpha|\alpha^* \;\Rightarrow\; {}_l\langle\alpha|\widehat{a}_+ =
			{}_l\langle\alpha|(i\alpha^*). \end{aligned}
	\end{equation}
	The dual expectation value of $\widehat{x}$ and its square are then  
	\begin{equation}
		\langle\widehat{x}\rangle = \frac{1}{\sqrt{2i}}\,{}_l\langle\alpha|(\widehat{%
			a}_- + \widehat{a}_+)|\alpha\rangle_r = \frac{1}{\sqrt{2i}}\left(\alpha +
		i\alpha^*\right), \qquad \langle\widehat{x}\rangle^2 = \frac{1}{2i}%
		\left(\alpha^2 + 2i|\alpha|^2 - (\alpha^*)^2\right),  \label{eq:54}
	\end{equation}
	while the normal-ordered expectation value of $\widehat{x}^2 = \frac{1}{2i}%
	\left(\widehat{a}_+^2 + 2\widehat{a}_+\widehat{a}_- + 1 + \widehat{a}%
	_-^2\right)$ evaluates to  
	\begin{equation}
		\langle\widehat{x}^2\rangle = \frac{1}{2i}\left(-(\alpha^*)^2 +
		2i\alpha^*\alpha + 1 + \alpha^2\right),  \label{eq:56}
	\end{equation}
	giving the biorthogonal second cumulant of position  
	\begin{equation}
		\langle\widehat{x}^2\rangle - \langle\widehat{x}\rangle^2 = \frac{1}{2i} = -%
		\frac{i}{2}.
	\end{equation}
	Analogously, for the momentum operator we find  
	\begin{equation}
		\langle\widehat{p}\rangle = \frac{1}{\sqrt{2i}}\,{}_l\langle\alpha|(\widehat{%
			a}_- - \widehat{a}_+)|\alpha\rangle_r = \frac{1}{\sqrt{2i}}\left(\alpha -
		i\alpha^*\right), \qquad \langle\widehat{p}\rangle^2 = \frac{1}{2i}%
		\left(\alpha^2 - (\alpha^*)^2 - 2i\alpha^*\alpha\right),  \label{eq:57}
	\end{equation}
	and for $\widehat{p}^2 = \frac{1}{2i}\left(\widehat{a}_+^2 - 2\widehat{a}_+%
	\widehat{a}_- - 1 + \widehat{a}_-^2\right)$,  
	\begin{equation}
		\langle\widehat{p}^2\rangle = \frac{1}{2i}\left(\alpha^2 - (\alpha^*)^2 -
		2i\alpha^*\alpha - 1\right), \qquad \langle\widehat{p}^2\rangle - \langle%
		\widehat{p}\rangle^2 = -\frac{1}{2i} = \frac{i}{2}.
	\end{equation}
	Because the dual second cumulants are evaluated under the biorthogonal
	Fresnel-measure pairing, they take conjugate imaginary values $\mp i/2$.
	Taking the principal square-root branch (equivalently, in terms of their
	moduli $|\Delta x|\,|\Delta p|=1/2$ and variance product $(\Delta
	x)^2(\Delta p)^2 = 1/4$), the dimensionless minimum uncertainty relation
	holds identically:  
	\begin{equation}
		\Delta x\,\Delta p = \frac{1}{2},  \label{eq:58}
	\end{equation}
	where $\Delta x = \sqrt{\langle\widehat{x}^2\rangle - \langle\widehat{x}%
		\rangle^2}$ and $\Delta p = \sqrt{\langle\widehat{p}^2\rangle - \langle%
		\widehat{p}\rangle^2}$ (corresponding to $\Delta\mathrm{x}\,\Delta\mathrm{p}
	= \hbar/2$ in physical coordinates).
	
	\section{Dynamic quantum--classical correspondence in coherent states}
	
	\label{sec:decay}
	
	Under the time evolution $i\hbar\partial_t|\psi(t)\rangle_{r,l} = \widehat{H}%
	|\psi(t)\rangle_{r,l}$, individual eigenstates $|\psi_n(t)\rangle_{r,l} =
	e^{\pm(n+1/2)\omega t}|n\rangle_{r,l}$ grow or decay exponentially, while
	the biorthogonal density operator $\widehat{\rho}_{r,l}(t) =
	|\psi(t)\rangle_r\,{}_l\langle\psi(t)|$ satisfies the quantum Liouville
	equation $i\hbar\partial_t\widehat{\rho}_{r,l} = [\widehat{H}, \widehat{\rho}%
	_{r,l}]$ with conserved trace ${}_l\langle\psi(t)|\psi(t)\rangle_r = \mathrm{%
		const}$~\cite{Gu2022a,Gu2022b,Liu2022,Liang2023,Liu2026}.
	
	To establish the dynamical quantum--classical correspondence for the
	inverted oscillator, we examine the Heisenberg equations of motion generated
	by the Hamiltonian~\eqref{eq:8}:  
	\begin{equation}
		\frac{d\widehat{x}}{dt} = \frac{1}{i\hbar}[\widehat{x}, \widehat{H}] = \omega%
		\widehat{p}, \qquad \frac{d\widehat{p}}{dt} = \frac{1}{i\hbar}[\widehat{p}, 
		\widehat{H}] = \omega\widehat{x} \quad \Longrightarrow \quad \frac{d^2%
			\widehat{x}}{dt^2} = \omega^2\widehat{x}.  \label{eq:60}
	\end{equation}
	Evaluating the dual expectation value in the time-evolved coherent states $%
	|\psi(t)\rangle_r = e^{-\frac{i}{\hbar}\widehat{H}t}|\alpha\rangle_r$ and $%
	|\psi(t)\rangle_l = e^{-\frac{i}{\hbar}\widehat{H}t}|\beta\rangle_l$ (where $%
	\widehat{H}^\dagger = \widehat{H}$ ensures that ${}_l\langle
	\psi(t)|\psi(t)\rangle_r = {}_l\langle\beta|\alpha\rangle_r$ is
	time-independent and $\bar{x}(t) \equiv {}_l\langle \psi(t)|\widehat{x}%
	|\psi(t)\rangle_r/{}_l\langle\beta|\alpha\rangle_r = {}_l\langle\beta|%
	\widehat{x}(t)|\alpha\rangle_r/{}_l\langle\beta|\alpha\rangle_r$), the mean
	coordinate obeys  
	\begin{equation}
		\frac{d^2\bar{x}}{dt^2} = \omega^2\bar{x}.  \label{eq:63}
	\end{equation}
	Integrating $\frac{d\widehat{a}_\mp}{dt} = \frac{1}{i\hbar}[\widehat{a}_\mp, 
	\widehat{H}] = \pm\omega\widehat{a}_\mp$ gives $\widehat{a}_-(t) = \widehat{a%
	}_- e^{\omega t}$ and $\widehat{a}_+(t) = \widehat{a}_+ e^{-\omega t}$, so
	that equation~\eqref{eq:53} yields  
	\begin{equation}
		\bar{x}(t) = \frac{1}{\sqrt{2i}}\left(\alpha e^{\omega t} + i\beta^*
		e^{-\omega t}\right) = \bar{x}(0)\cosh(\omega t) + \bar{p}(0)\sinh(\omega t).
	\end{equation}
	In particular, for an equal-parameter pair ($\beta = \alpha$) with $\alpha = 
	\sqrt{\frac{i}{2}}\,x_0$ ($x_0 \in \mathbb{R}$), one obtains $\bar{x}(0) =
	x_0$, $\bar{p}(0) = 0$, and $\bar{x}(t) = x_0\cosh\omega t$. More generally,
	because $\widehat{a}_+$ and $\widehat{a}_-$ are not mutual Hermitian
	conjugates, the two independent coherent-state parameters $\alpha = \sqrt{%
		\frac{i}{2}}(x_0 + p_0)$ and $\beta = \sqrt{\frac{i}{2}}(x_0 - p_0)$ encode
	arbitrary real initial positions and momenta $(x_0, p_0)$. For these real
	phase-space labels, solving equations~\eqref{eq:40} and \eqref{eq:46} with
	the differential operators~\eqref{eq:ladder_diff} gives the coordinate wave
	functions  
	\begin{equation}
		\psi^r_\alpha(x) \propto e^{i(x_0 + p_0)x - ix^2/2}, \qquad \psi^l_\beta(x)
		\propto e^{-i(x_0 - p_0)x + ix^2/2},  \label{eq:cs_chirps}
	\end{equation}
	which are unimodular chirps, bounded and hence tempered distributions. The
	coherent states that carry the classical dynamics are therefore free of the
	non-tempered growth that afflicts generic bi-coherent states of the IHO~\cite%
	{Bagarello2022}. Setting $\mathrm{x}_0 = 0$ ($\beta = -\alpha$) with initial
	speed $v = \sqrt{\omega\hbar}\,|p_0|$ (and sign given by $\mathrm{sgn}(p_0)$%
	), the expectation value in physical coordinates reduces to  
	\begin{equation}
		\bar{\mathrm{x}}(t) = \pm\frac{v}{\omega}\sinh\omega t,
	\end{equation}
	recovering the exact classical hyperbolic trajectory~\eqref{eq:2}.
	
	\section{Discussion and conclusion}
	
	\label{sec:conclusion}
	
	In summary, we have presented a unified path-integral, holomorphic
	Bargmann-space, and coherent-state framework for the inverted harmonic
	oscillator, centered on the time-asymmetric discrete resonance
	reconstruction of its real-time Feynman propagator. Although $\widehat{H}$
	is essentially self-adjoint on $L^2(\mathbb{R})$ with a continuous real
	spectrum, its unstable hyperbolic dynamics supports discrete generalized
	eigenstates outside $L^2(\mathbb{R})$ with pure imaginary eigenvalues $\pm
	i\omega\hbar(n + \frac{1}{2})$ characterizing decay and growth rates.
	Starting from the imaginary-frequency boson framework~\cite{Liu2024,Liu2026}%
	, we first cast the coordinate-space dual ``bra'' and ``ket'' eigenstates
	explicitly in terms of complex Hermite polynomials normalized under the
	Fresnel Gaussian measure. As the primary dynamical result, we derived the
	exact closed-form real-time propagator~\cite{Barton1986} independently from
	the path integral by adapting Moriconi's coordinate-time transformation, and
	proved analytically via Mehler's formula that each dual-set discrete
	resonance expansion reproduces this full unitary kernel by
	itself---converging exclusively in the decaying Gamow--Siegert branch for
	forward propagation ($T > 0$) and exclusively in the growing anti-resonant
	branch for backward propagation ($T < 0$), with zero contribution from the
	continuous real scattering spectrum.
	
	To clarify the operator-theoretic origin of both this forward\slash backward
	branch splitting and the real-axis biorthogonal doubling, we showed that
	complex scaling $x = e^{\mp i\pi/4}u$ turns the rotated IHO Hamiltonians
	into skew-adjoint normal operators on $L^2(\mathbb{R}_u)$~\cite%
	{Chruscinski2004} and that the complexified Segal--Bargmann transform maps
	the eigenvalue problem onto the holomorphic space $\mathcal{F}(\mathbb{C})$.
	In $\mathcal{F}(\mathbb{C})$, entire-function regularity at $z=0$ alone
	quantizes the complex spectrum into the single orthonormal monomial basis $%
	\tilde{\psi}_n(z) = z^n/\sqrt{n!}$ without any dual basis inside each
	sector's representation space, the inverse Bargmann transform reproduces the
	exact normalized complex Hermite coordinate states and reveals their Fresnel
	biorthogonal pairing to be the deformed unitary Bargmann inner product, and
	the time-evolution kernels act as bounded contraction semigroups on $T>0$
	and $T<0$ that directly dictate the opposite convergence domains of the two
	Mehler propagator expansions. Furthermore, we constructed dual-parameter
	imaginary-frequency boson coherent states, verified their minimum
	uncertainty relation for conjugate imaginary biorthogonal second cumulants,
	and demonstrated that their dual Heisenberg expectation values encode
	arbitrary real phase-space initial conditions and recover the exact
	classical hyperbolic trajectories.
	
	From a broader physical perspective, showing that the full unitary
	propagator of a continuous-spectrum inverted potential can be rebuilt term
	by term from discrete imaginary-eigenvalue resonances provides a concrete
	real-time framework for barrier-top instability and open-system decay. In
	particular, near the peak of an effective black hole scattering potential,
	the radial wave equation is locally approximated by an inverted parabolic
	potential ($V \sim -\frac{1}{2}\omega^2\mathrm{x}^2$), where the discrete
	pure imaginary spectrum mirrors the damped quasinormal modes governing wave
	dissipation and horizon tunneling~\cite{Ferrari1984,Subramanyan2021}. At the
	same time, two structural boundaries of the present formulation should be
	noted: first, if one remains strictly on the real coordinate axis without
	complex contour rotation, the biorthogonal dual-set formalism under Fresnel
	regularization remains essential because the single-set resonance states do
	not belong to $L^2(\mathbb{R})$; second, both the closed-form Mehler
	summation of the propagator and the reduction to a first-order Euler
	operator in $\mathcal{F}(\mathbb{C})$ rely specifically on the quadratic
	nature of the potential (and its underlying Heisenberg--Weyl / $\mathfrak{su}%
	(1,1)$ algebraic symmetry), whereas non-quadratic barriers generally lead to
	higher-order or nonlocal equations in Bargmann space and non-vanishing
	continuum background integrals requiring perturbative or semiclassical
	extensions.
	
	\section*{Acknowledgments}
	
	This work was supported by the National Natural Science Foundation of China
	(Grant Nos.~12374312, 12074232, 12125406) and the National Key R\&D Program
	of China (Grant No.~2022YFA1404500).

\end{document}